# Magnetic susceptibility, phonons and dielectric constant of single crystalline $BiFeO_3$

**Jun Lu**[1,2], **M Schmidt**[1], **P Lunkenheimer**[1], **A Pimenov**[3], **A A Mukhin**[4], **V D Travkin**[4] **and A Loidl**[1]

[1] Experimental Physics V, Center for Electronic Correlations and Magnetism, University of Augsburg, 86135 Augsburg, Germany
[2] School of Materials Science and Engineering, University of Science and Technology, Beijing 100083, China
[3] Experimentelle Physik IV, Universität Würzburg, 97074 Würzburg, Germany
[4] Prokhorov General Physics Institute of the Russian Academy of Sciences, 119991 Moscow, Russia

E-mail: lujun@mater.ustb.edu.cn

**Abstract.** The magnetic susceptibility between 1.5 and 800 K, the infrared reflectivity at room temperature and the temperature dependence of the dielectric constant at mm-wavelengths are reported for single crystalline $BiFeO_3$. A well developed anomaly in the magnetic susceptibility signals the onset of antiferromagnetic order close to 640 K. Beside this anomaly, no further indications of phase or spin-glass transitions are found in the susceptibility down to the lowest temperatures. From infrared reflectivity we were able to identify all 9 phonon modes which are expected to be infrared active within the **ab** plane of the crystal. The temperature dependence of the dielectric constant was measured contact free in the 100 GHz range yielding $\varepsilon' = 54$ at room temperature. The loss is substantial and strongly frequency dependent indicating the predominance of hopping conductivity.

After extensive work in the late sixties [1], multiferroics have gained considerable attention in recent years driven by the progress that has been achieved in sample preparation, measuring techniques and theoretical concepts [2,3]. Multiferroics promise important applications because in the case of coupled ferroelectric and magnetic order parameters the magnetization can be switched by electric and the polarization by magnetic fields enabling, e.g., significant progress in data storage processes. $BiFeO_3$ (BFO) is a rhombohedrally distorted perovskite with space group R3c and belongs to the rare class of materials with long range magnetic and long range polar order already at room temperature [4]. However, BFO is extremely hard to synthesize in pure and stoichiometric form and many experiments suffer from impurity phases, magnetic defect ions or deviations from the ideal oxygen stoichiometry.

Despite these drawbacks, it seems nowadays established that BFO undergoes polar order close to 1100 K [5] and becomes antiferromagnetic at 630 K. Antiferromagnetic order with a G-type spin configuration was established by Kiselev *et al.* [6]. Later on, high-resolution neutron scattering revealed that the collinear G-type structure is modified by a long-range modulation and that the ground state of BFO can best be represented as cycloidal spiral with a period of approximately 62 nm [7]. Utilizing the same single crystal, ac susceptibility measurements revealed a magnetic anomaly at zero magnetic field close to 400 K, interpreted as a change in the anisotropy energy [8]. Since then, numerous experimental reports have pointed towards further possible magnetic phase or spin-glass transitions at lower temperatures [9,10], which, however, do not provide a conclusive picture.

Similar discrepancies exist concerning the phonon properties. There are a number of Raman [11,12,13] and infrared (IR) experiments [14,15] on BFO including first principle calculations [16]. Group theory predicts 9 E modes which are IR active within the **ab** plane and 4 $A_1$ modes which are IR active along the **c** axis. However, the so far published results neither agree concerning the eigenfrequencies nor concerning the assignment of the modes. A Raman work focusing on low-frequency electromagnons [17] has been performed by Cazayous et al. [18]. Finally, as already at room temperature ac conductivity dominates the dielectric response in low-frequency ac measurements (< 1 GHz), no reliable data of the dielectric constant at room temperature exist [14].

In this short communication we address all these relevant questions: Via magnetic measurements on a high quality single crystal we show that no low-temperature ($T < T_N$) phase or spin-glass transitions exist in BFO. Using the same high-quality single crystal we determine all 9 IR active phonon modes in the **ab** plane and can assign them unambiguously. Finally, we determine the exact value of the dielectric constants for temperatures between 80 K and 300 K.

The BFO single crystal (5 × 4 × 0.4 mm) was grown by flux methods as described elsewhere [19]. The magnetic measurements were performed utilizing a SQUID magnetometer. For the reflectivity measurements a Bruker Fourier-transform spectrometer IFS 113v was used. Finally, the complex dielectric permittivity was determined from measurements of transmittance and phase shift in the millimetre-wave range employing a THz spectrometer utilizing the Mach-Zehnder geometry [20].

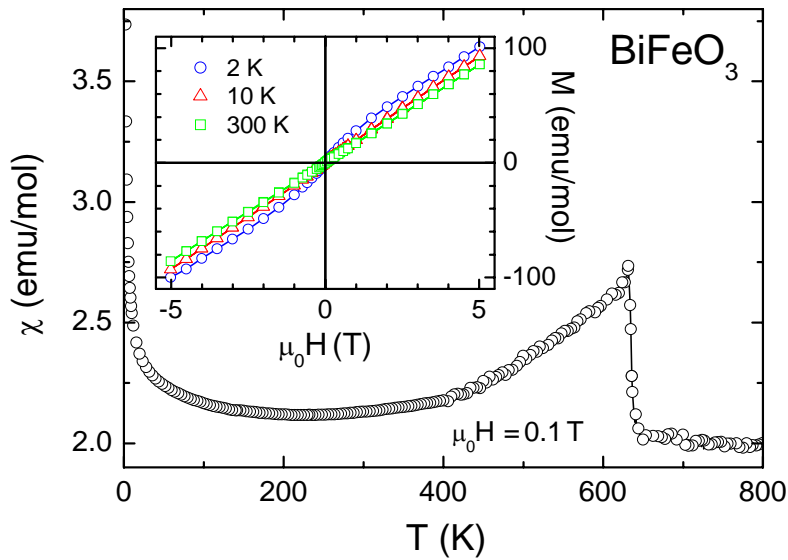

**Figure 1.** Magnetic susceptibility vs. temperature of $BiFeO_3$ as measured in an external magnetic field of 0.1 T. For these measurements field-cooled conditions have been chosen. The inset shows the magnetization vs. magnetic field at 2, 10 and 300 K.

Figure 1 shows the magnetic susceptibility of BFO from 1.5 K to 800 K. A clear anomaly close to 640 K signals the onset of long-range antiferromagnetic order. The anomaly looks rather untypical for an antiferromagnet with an anomalous continuous decrease of the susceptibility below $T_N$ arriving at a room temperature value of 2.2 emu/mol. The impurity-induced Curie-type increase towards 0 K is of extremely low amplitude signalling an almost impurity-free single crystal. Our results at low temperatures roughly compare with reports by Lebeugle et al. [21], but reveal an even lower susceptibility indicative for a high purity sample. In this pure crystal no anomalies can be detected which might indicate further magnetic phase transitions. We also performed field cooled and zero-field cooled susceptibility measurements including ac experiments at low fields and found no indications for any type of spin-glass ordering, contrary to reports in [9]. The inset shows magnetization measurements at 2, 10 and 300 K indicating the absence of ferromagnetic hystereses.

Figure 2 shows the reflectivity between 50 and 650 cm$^{-1}$. The measurements have been performed with the electrical field parallel to the **ab** plane. We also tried polarized experiments varying the field direction within the **ab** plane, but could detect almost no anisotropy. The reflectivity clearly indicates 9 modes. The solid line in figure 2 represents a standard three-parameter fit of the reflectivity for each mode utilizing transverse optical eigenfrequencies, the dielectric strength and the inverse lifetime as fit parameters. In these calculations a high-frequency dielectric constant $\varepsilon_\infty = 5.4$ was used. Due to the rather restricted high-frequency limit of the performed measurements, this value is not very reliable.

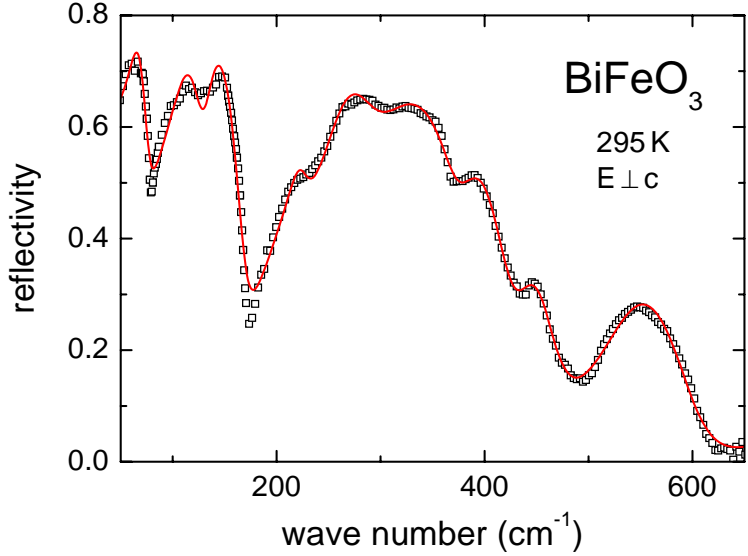

**Figure 2.** Reflectivity vs. wave number of BiFeO$_3$. In these measurements the electrical field was parallel to the **ab** plane, an experimental setup sensitive to E modes. The solid line represents the result of a three-parameter fit for each mode.

The results of these fits are indicated in table 1. Due to the high quality of the reflectivity data, the observation and correct assignment of all 9 IR active E modes is of high significance. Remarkably, they are neither fully consistent with previous experimental reports [14,15], nor with first principle calculations [16]. Specifically, the model calculations miss the first eigenmode close to 64 cm$^{-1}$ and predict a mode at 274 cm$^{-1}$, which according to our interpretation is missing. However, we admit that small contributions to the reflectivity appear close to 100 and 280 cm$^{-1}$. If these are leakage modes from other symmetry directions or multiphonon processes is unclear. Table 1 documents that the damping of the modes is strong, sometimes exceeding 20% of the eigenfrequency. It is also interesting that the dielectric strength $\Delta\varepsilon$ mainly results from the first two modes. The sum over all modes yields $\Delta\varepsilon = 49.6$, leading to a static dielectric constant $\varepsilon' = 55$ at room temperature. Our value is enhanced when compared to the reflectivity measurements on polycrystalline samples with $\varepsilon' = 30$ [14], but is significantly lower than the value obtained on single crystals with $\varepsilon' = 83$ [15].

**Table 1.** Eigenfrequency $\nu$, dielectric strength $\Delta\varepsilon$ and damping $\gamma$ of the 9 E modes in BiFeO$_3$.

| Mode | $\nu$ (cm$^{-1}$) | $\Delta\varepsilon$ | $\gamma$ (cm$^{-1}$) |
|---|---|---|---|
| E(1) | 64.1 | 21.1 | 13.5 |
| E(2) | 110.7 | 14.3 | 25.4 |
| E(3) | 137.3 | 4.23 | 19.3 |
| E(4) | 221.7 | 1.42 | 22.7 |
| E(5) | 260.2 | 5.38 | 42.8 |
| E(6) | 309.5 | 2.16 | 73.1 |
| E(7) | 383.0 | 0.27 | 43.4 |
| E(8) | 440.7 | 0.17 | 41.3 |
| E(9) | 535.7 | 0.56 | 76.4 |

Finally, figure 3(a) shows $\varepsilon'(T)$ between 80 K and 300 K, measured at 4 cm$^{-1}$. We found a room temperature value of $\varepsilon'$ close to 53, which decreases with decreasing temperature. These values, which agree well with the IR results, should be very close to the intrinsic $\varepsilon'$. In addition, we determined the dielectric loss $\varepsilon''$ [figure 3(b)], which also decreases during cooling and strongly depends on frequency. The observed linear increase with frequency does not match the extrapolated tail of the lowest phonon mode. $\varepsilon'' \propto \nu$ implies a quadratic increase of the conductivity, $\sigma' \propto \nu^2$. This may indicate Mott-like hopping conduction as considered, e.g., in [22].

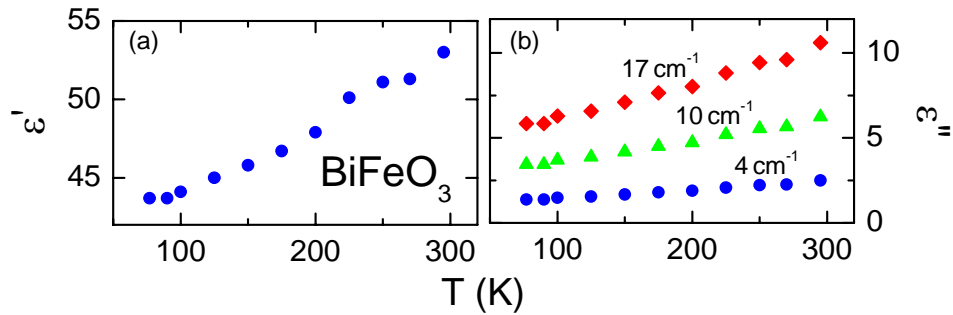

**Figure 3.** $\varepsilon'(T)$ (a) and $\varepsilon''(T)$ (b) of BiFeO$_3$ at various frequencies.

We thank V.A. Murashov for providing BFO crystals. This work was supported by the DFG via the SFB 484. J. Lu acknowledges support from China Scholarship Council for his oversea Ph.D. studies.